\documentclass[a4paper,12pt]{article}

\usepackage{graphicx}
\usepackage{subfigure} 
\usepackage[UKenglish]{babel} 
\usepackage{xspace} 
\usepackage{amsbsy} 
\usepackage[version=3]{mhchem}

\usepackage[left=2.5cm,right=2.5cm,top=2cm,bottom=2cm]{geometry}

\usepackage{fancyhdr}
\fancypagestyle{plain}{ %
\fancyhf{} 
\lhead{\textit{Precipitate assemblies in Al-Ag-Cu.}}
\rhead{J. M. Rosalie, L. Bourgeois \& B. C. Muddle}
\lfoot{\textit{Pre-print}}
\rfoot{\thepage}
}

\newcommand{\gammaprime}{\ensuremath{\gamma^{\prime}}\xspace}
\newcommand{\thetaprime}{\ensuremath{\theta^{\prime}}\xspace}
\newcommand{\thetadoubleprime}{\ensuremath{\theta^{\prime\prime}}\xspace}

\newcommand{\axis}[2][\alpha]{\ensuremath{\mathrm{[#2]_{#1}}}\xspace} 
\newcommand{\axes}[2][\alpha]{\ensuremath{\mathrm{\langle #2 \rangle_{#1}}}\xspace} 
\newcommand{\plane}[2][\alpha]{\ensuremath{\mathrm{(#2)_{#1}}}\xspace} 
\newcommand{\planes}[2][\alpha]{\ensuremath{\mathrm{\{#2\}_{#1}}}\xspace} 

\newcommand{\stdfig}[1]{\resizebox{6.5cm}{!}{\includegraphics{#1}}} 
\newcommand{\medfig}[1]{\resizebox{8.67cm}{!}{\includegraphics{#1}}} 

\newcommand{\burgers}[3]{\ensuremath{\boldsymbol{b}=\frac{#1}{#2}\langle #3 \rangle}\xspace} 

\newcommand{\alagA}{\ensuremath{\mathrm{Al_{98.3}Ag_{1.7}}}\xspace}
\newcommand{\alagB}{\ensuremath{\mathrm{Al_{99.2}Ag_{0.8}}}\xspace}
\newcommand{\alagcuB}{\ensuremath{\mathrm{Al_{98.2}Ag_{0.9}Cu_{0.9}}}\xspace}  
\newcommand{\alcuA}{\ensuremath{\mathrm{Al_{98.3}Cu_{1.7}}}\xspace}

\begin{document}
\title{Precipitate assemblies formed on dislocation loops in aluminium-silver-copper alloys}
\author{Julian~M.~Rosalie$^{a,b,c,\dagger}$
\and Laure~Bourgeois$^{b,c,d}$ \and
Barrington~C.~Muddle$^{b,c}$}
\date{}

\maketitle
\noindent
$^a$Microsctructure Design Group, Structural Metals Center, National Institute for Materials Science (NIMS), Japan.\\
$^b$ARC Centre of Excellence for Design in Light Metals, Australia.\\
$^c$Department of Materials Engineering, Monash University, 3800, Victoria, Australia.\\
$^d$Monash Centre for Electron Microscopy, Monash University, 3800, Victoria, Australia.\\
$^\dagger$Present address, work carried out at $\mathrm{^c}$.

\begin{abstract}

The precipitation microstructure of the \gammaprime (\ce{AlAg2})
intermetallic phase has been examined in aluminium-silver-copper alloys. 
The microstructure developed in an Al-0.90at.\%Ag-0.90at.\%Cu alloy was significantly different from that reported for binary Al-Ag alloys. 
The orientation relationship between the matrix and precipitate was unchanged;
however, the \gammaprime phase formed as aggregates with a two-dimensional open assemblies.  
Each such assembly contained two variants of the \gammaprime phase alternately arranged to form a faceted elliptical unit.
The \thetaprime (\ce{Al2Cu}) 
phase formed on these assemblies after further ageing.
The faceted elliptical assembly morphology has not been previously reported for the \gammaprime precipitate. 
The change in precipitation behaviour was attributed to copper modifying the as-quenched defect structure of the matrix.
This precipitation morphology clarifies earlier observations on the precipitate number density and mechanical properties  of aluminium-silver-copper alloys.
\end{abstract}

\section*{Introduction and background} 

Aluminium-silver  alloys have been frequently selected to investigate the solid-state phase transformations involved in precipitation strengthening\cite{passoja:1971, shchegoleva:1976, shchegoleva:1981,howe:1985, howe:1985a, howe:1987, muddle:1994, nie:1999}.
The system is not of commercial interest, 
owing to the high cost (and density) of silver and the moderate precipitation strengthening response.
However, 
it is an exceptionally useful system to study precipitation of intermetallic phases because the principle strengthening phase, 
the \gammaprime (\ce{AlAg2}) 
intermetallic precipitate, 
is formed with negligible \((-2.5\%)\) volumetric strain\cite{muddle:1994a}.  

The nucleation of the \gammaprime (\ce{AlAg2}) 
phase is considered difficult despite the coherent precipitate-matrix interface, 
low interfacial energy\cite{ramanujan:1992} and the  minimal volumetric strain\cite{muddle:1994a} associated with the phase transformation.  
Silver-rich Guinier-Preston (GP) zones, 
whose formation precedes that of \gammaprime,
have been observed after 200\,h ageing at 175$^\circ$C\cite{borchers:1969a}, indicating that significant amounts of silver remained partitioned to the GP zones, rather than precipitating as the \gammaprime intermetallic phase. 
Precipitation of \gammaprime precipitates generally requires the existence of suitable heterogeneous sites such as grain boundaries\cite{clark:1967}, dislocations \cite{frank:1961, nicholson:1961, passoja:1971} or dislocation loops\cite{frank:1961} to assist in nucleation. 
The \gammaprime phase is thought to form via the diffusion of silver to a pre-existing stacking fault, 
and the simultaneous extension of the stacking fault due to the lower stacking fault energy of silver\cite{voss:1999}.

The Al-Ag-Cu phase diagram contains only binary phases in the aluminium-rich region \cite{liu:1983} and the equilibrium phases identified are those found in the Al-Ag and Al-Cu binary systems.
The \gammaprime phase has also been reported in Al-Ag-Cu  alloys\cite{bouvy:1965,borchers:1969},  
along with copper-containing \thetadoubleprime (\(\mathrm{Al_3Cu}\))  
and \thetaprime (\ce{Al2Cu})
intermediate phases\cite{bouvy:1965,khatanova:1966,borchers:1969a,guinier:1942, konno:2006}. 
The equilibrium $\gamma$ (\ce{AlAg2}) 
and $\theta$ (\ce{Al2Cu})
phase were also found after high temperature ageing\cite{bouvy:1965}.

In aluminium-rich Al-Ag and Al-Ag-Cu alloys 
the \gammaprime phase formed thin, 
high aspect ratio plates, 
with \planes[]{111} habit. 
The \thetaprime (\ce{Al2Cu})
phase also retains the same plate-like morphology 
and \planes[]{100} habit in both Al-Cu and Al-Ag-Cu alloys\cite{guinier:1942, konno:2006}. 
The orientation relationships determined for \gammaprime and \thetaprime in Al-Ag-Cu ternary alloys 
\cite{zakharova:1966} were identical to those observed in the respective binary alloys \cite{muddle:1994a}. 

Considerable differences in the precipitation of the \gammaprime phase in the ternary alloys. have been reported\cite{borchers:1969a,borchers:1969b,khatanova:1966,rosalie:2005}. 
The \gammaprime plates in ternary alloys were found to be finer in scale and were formed in greater number densities than in binary Al-Ag \cite{borchers:1969a}.
Recent quantitative measurements found \gammaprime number density were an order of magnitude greater in the ternary alloy compared to binary alloys aged under identical conditions\cite{rosalie:2005}.
The \gammaprime precipitates also grew significantly faster than \thetaprime~(\ce{Al2Cu}) 
precipitates in  Al-7wt\%Ag-3wt\%Cu alloys\cite{khatanova:1966}.

The microstructural changes in  the ternary alloy have been attributed to competition between copper and silver for vacancies\cite{borchers:1969a};
however, no detailed mechanism has been proposed and the role of  copper is unclear.
Copper generates considerable misfit strains in the aluminium matrix  (unlike silver) and is known to modify the defect structure in binary aluminium-copper alloys\cite{thomas:1959,westmacott:1971}. 
These changes include changes in dislocation loop geometry and the formation of non-faulted loops \cite{westmacott:1971}. 

The higher number density and more refined \gammaprime precipitate distribution 
\cite{borchers:1969a,rosalie:2005} suggest that the nucleation of the \gammaprime phase may be enhanced in the ternary alloys. 
This is of interest given the difficulty in precipitating the \gammaprime phase;
however,
a detailed characterisation of the microstructure of aluminium-silver-copper alloys is lacking. 
Such an investigation into the ternary Al-Ag-Cu system is warranted in order to determine how microstructure differs from that of the binary alloys and how the addition of copper leads to a finer microstructure.
Means of increasing the precipitate number density are of interest in alloy design and offer the potential of increasing the effectiveness of precipitation strengthening\cite{zhu:2002}.  

The present work sets out to determine the reasons behind the microstructural refinement of the \gammaprime phase in Al-Ag-Cu alloys. 
This involves an examination of the microstructure of a ternary Al-Ag-Cu using conventional and high resolution transmission electron microscopy (TEM and HRTEM). 

\section*{Experimental details}

The ternary aluminium-silver-copper alloy used in this work was cast from high-purity aluminium (Cerac alloys, 99.99\% purity), 
silver (AMAC alloys, 99.9+\%) and copper (AMAC alloys, 99.99\%). 
The composition used contained 0.9at.\%Ag and 0.90at.\%Cu, hereafter designated \alagcuB. 
Aluminium-copper and alumin\-ium-silver alloys were also cast to provide a basis for comparison.
The binary compositions were selected so that the level of alloying elements was a) approximately equal to the combined Ag and Cu content of the ternary alloy or b) approximately equal to the Ag content of the ternary alloy. 
The pure elements were melted in air at 700$^\circ$C in a graphite crucible, stirred, 
then poured into graphite-coated steel molds. 
The cast ingots were homogenised at 525$^\circ$C for 7\,days and hot-rolled to 2\,mm (for hardness testing), 
or 0.5\,mm for further thinning and TEM analysis.
Alloy compositions were confirmed by inductively coupled plasma atomic emission spectrometry (ICP-AES). 
The principle impurities detected (Fe and Si) were present at less that 0.03 at.\%. 

Solution treatments were carried out for 0.5\,h at 525$^\circ$C in a nitrate/nitrite salt pot.
Samples were quenched into water at ambient temperature and aged in an oil bath at 200$^\circ$C for up to 72\,h.

Vickers hardness testing was carried out using loads of 2.5\,kg--10\,kg, depending on the hardness of the material. 
The consistency of results obtained at each load was compared, 
and the difference was roughly $\mathrm{\pm1}$ Vickers hardness number (VHN).

Foils for TEM analysis were jet electro-polished 
at -20$^\circ$C using a nitric acid (33\%v/v) / methanol(67\%v/v) solution. 
The voltage applied was $-13$\,V, 
and currents averaged 200\,mA. 

Conventional transmission electron microscopy investigations were carried out using a Philips CM20 transmission electron microscope operating at 200\,kV. 
High resolution studies made use of a JEOL~2011 HRTEM instrument with a point resolution of 0.23\,nm.

\section*{Results}

\subsection*{Hardness measurements}

\begin{figure}
\begin{center}
\medfig{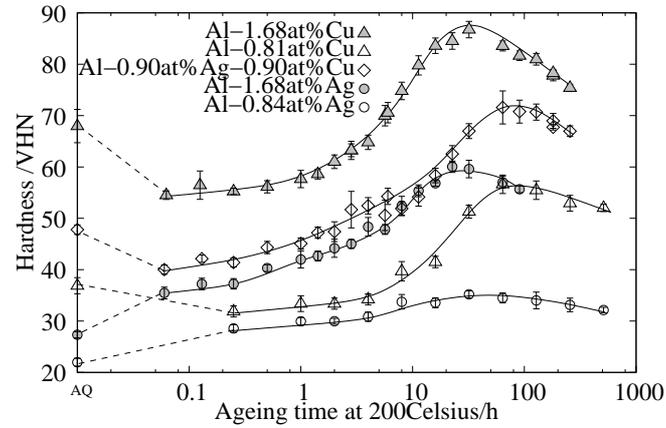} 
\caption[Hardness curves for Al-Ag and Al-Ag-Cu alloys.]
{Hardness curves (VHN) for Al-Ag alloys  aged at 200$^\circ$C.  
The curves are intended as a guide for the eye only. 
\label{fig1}}
\end{center}
\end{figure}

The Vickers hardness curves measured for the ternary \alagcuB alloy are presented in Figure~\ref{fig1},
along with those for Al-Ag\cite{Rosalie2008a} and Al-Cu\cite{rosalie:2005} binary  alloys. 
The copper containing alloys experienced a more significant increase in hardness than those with equal atomic amounts of silver. 
The response of the \alagcuB alloy was moderate, 
reaching a peak hardness of $\mathrm{72 \pm 3}$\,VHN after 64\,h of ageing.
This was greater than the peak hardness for a binary Al-Cu alloy with equivalent Cu (\alagB) but less than of \alcuA. 
The latter alloy had a total solute element level close to that of the alloy of interest, meaning that on a atomic percentage basis copper was a more effective strengthening element than silver.
The \alagcuB alloy required more prolonged ageing to reach maximum hardness (64\,h) than the \alcuA, 
\alagA (32\,h) and \alagB alloys (32\,h). 

\subsection*{Quenched-in defect structure \label{para-1to1-aq}}
\begin{figure}
\begin{center}
\stdfig{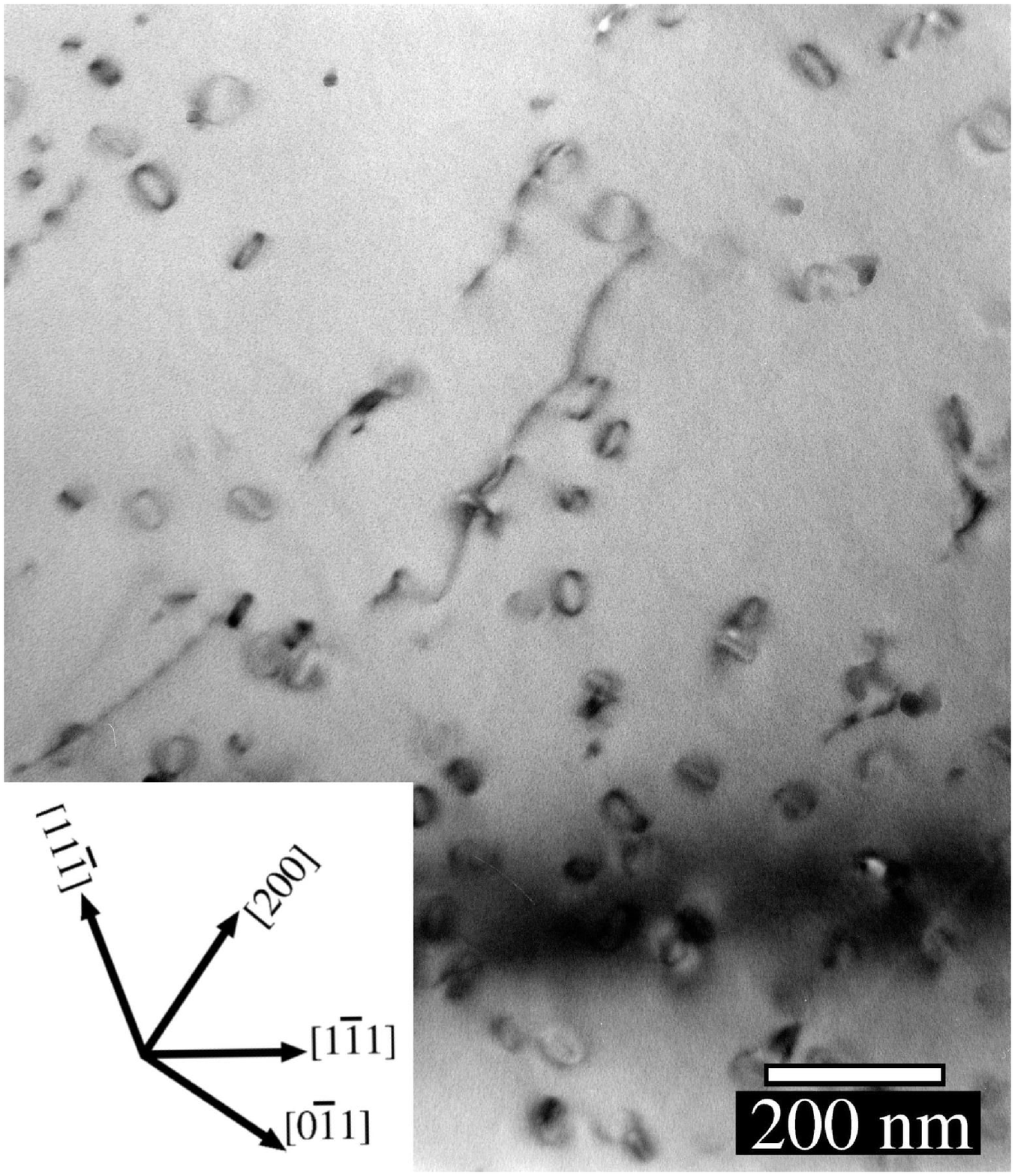} 
\caption{The as-quenched defect structure of the \alagcuB alloy,  
viewed close to the  \axis{011} zone in a two-beam condition with 
$\boldsymbol{g}=\mathrm{11\overline{1}}$.
The microstructure contains a population of elliptical dislocation loops and some linear dislocations. 
 \label{fig2}}
\end{center}
\end{figure}

The quenched-in defect structure was analysed using two-beam diffraction contrast TEM. 
Figure~\ref{fig2} 
shows an electron micrograph of the Al-Ag-Cu alloy in the as-quenched condition. 
 Elongated elliptical dislocation loops are clearly visible on the \planes[]{110} planes of the \(\alpha\)-aluminium matrix. 
 Stacking fault contrast would appear as diffuse bands of alternate bright-dark contrast\cite{whelan:1957} and was not observed for the dislocation loops in this alloy.
These dislocation loops were identified as perfect dislocation loops with
\burgers{1}{2}{110}, 
as expected for an aluminium-copper alloy with levels of solute around 0.90at.\%
\cite{westmacott:1971}.

\subsection*{Precipitate microstructure\label{para-1to1-prec}}

\begin{figure}
\begin{center}
\resizebox{7.5cm}{!}{\includegraphics{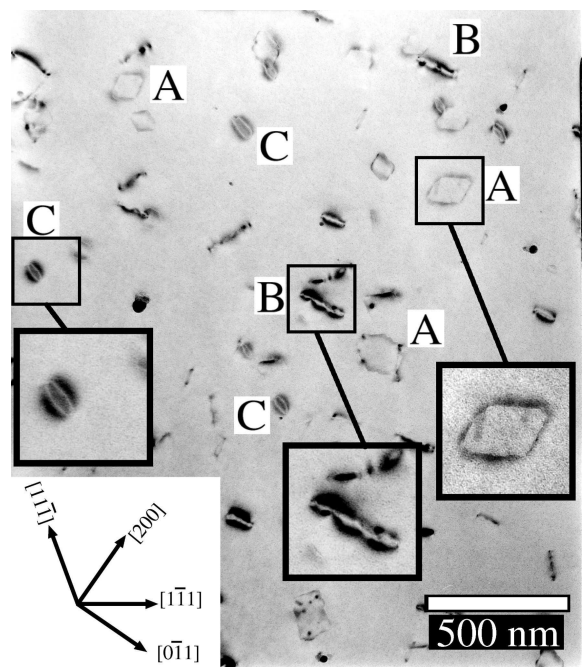}} 
\caption{Electron micrograph providing an overview of the microstructure of the \alagcuB alloy in the early stages of ageing. 
The microstructure contains elliptical assemblies of \gammaprime precipitates in which each precipitate is normal to the viewing direction ($A$), 
assemblies of precipitates where the plates appear to be inclined to the viewing direction ($B$), 
and dislocation loops ($C$). 
The loops appear here with inside contrast. 
The image was obtained parallel to the \axis{011} axis of a foil aged for 0.16\,h at 200$^\circ$C.
\label{fig3}}
\end{center}
\end{figure}

An overview of the microstructure for an ageing time of 0.16\,h at 200$^\circ$C is presented in Figure~\ref{fig3}. 
The microstructure contains what will be shown to be aggregated \gammaprime precipitates (at points labelled $A$ and $B$). 
Some of these aggregates (e.g. $A$) are comprised of two variants of the \gammaprime precipitate, 
with both variants normal to the beam direction.
This yields a two-dimensional assembly structure resembling a faceted ellipse, 
with the assembly lying normal to the \axis{011} and elongated parallel to a 
\axis{200} direction (see Figure~\ref{fig3})
Precipitate assemblies ($B$) appeared with the loop normal perpendicular to the viewing direction. 
In this orientation the \gammaprime precipitates are inclined to the beam and in poor contrast. 
Dislocation loops (labelled $C$) are also present and have inside contrast in this figure. 

\begin{figure}
\begin{center}
\subfigure[\label{fig4a}]{\stdfig{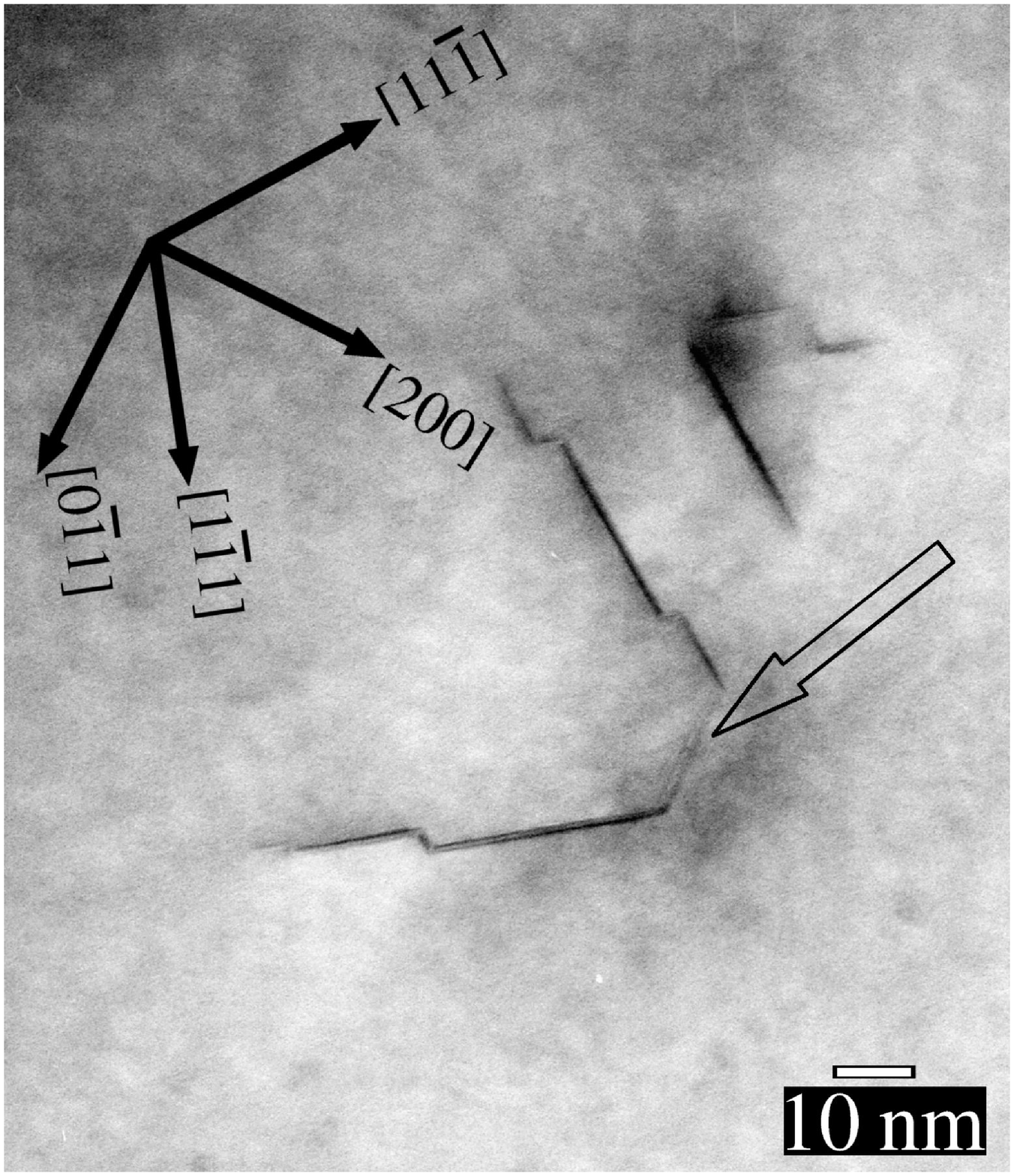}} 
\hfill
\subfigure[\label{fig4b}]{\stdfig{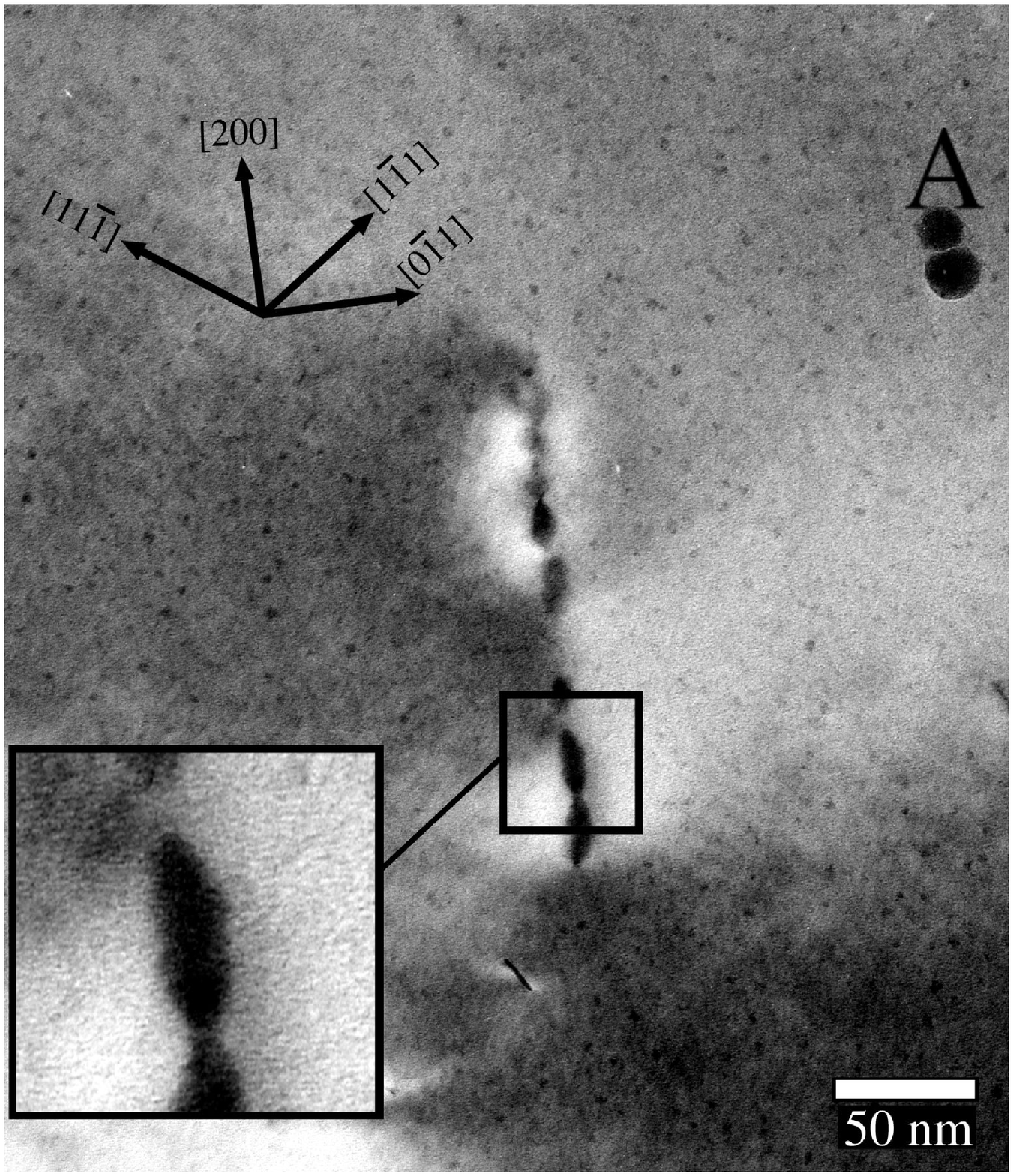}} 
\caption{The microstructure of the \alagcuB alloy after ageing for 0.083\,h  at 200$^\circ$C, 
viewed parallel to the \axis{011} zone.
In (a) a portion of a precipitate assembly and constituent precipitates are perpendicular to the viewing direction, 
while in (b) the assembly is viewed edge-on. 
In (a) the remainder of the assembly is absent due to foil trucncation. 
Note that \gammaprime precipitates have not formed at the end of the faceted loop , 
although a weaker, 
diffuse contrast features is present (arrowed).
In (b), 
note that the precipitates have grown further in the plane of the precipitate assembly.
Guinier-Preston zones are visible as the small, darker spheroids in (b).
The larger, more strongly absorbing spheroids at point $A$ in (b) are electro-deposited silver.
\label{fig4}}
\end{center}
\end{figure}

The structure of the elliptical precipitate assemblies is shown in greater detail in Figure~\ref{fig4}. 
The micrographs were obtained from a foil aged for 0.083\,h at 200$^\circ$C.
The precipitates plates are present as aggregates with alternating \planes{111} variants formed adjacent to one another, as shown in Figure~\ref{fig4a} 
which shows a portion of a precipitate assembly. 
(The remainder of this assembly was removed due to foil truncation.) 
Individual assemblies varied considerably in terms of the number of \gammaprime precipitates present,
their projected diameters  
and the size and shape of the ellipse. 
Precipitation of \gammaprime rarely occurred in the portion of the ellipse  parallel to the traces of \plane{200} planes
(see Figure~\ref{fig4a} ), 
leaving the assemblies as open, crown-like, two-dimensional structures. 
A diffuse feature is visible (arrowed), but the weaker, less defined contrast suggests that this is associated with strain or a dislocation rather than a precipitate.

Figure~\ref{fig4b}  shows an image obtained along the \axis{011} zone axis in which the assembly was viewed edge-on. 
While the individual \gammaprime precipitates possessed the standard  \planes{111} habit for the phase, 
the assembly as a whole was parallel to the  \planes{0\overline{1}1} plane. 
In this orientation the precipitates were seen in a line with the loop only partially occupied. 
Precipitate lengthening appeared most extensive along the habit plane of the assembly. 

Longer ageing times resulted in precipitates with greater diameter, 
and in more frequent impingement between adjacent precipitates, 
The  \gammaprime precipitates formed a continuous circuit after 0.5\,h of ageing
except at the portions tangential to the \planes{200} traces, which were devoid of \gammaprime precipitates.
The diffuse contrast feature noted in Figure~\ref{fig4a} was frequently observed in this region. 

\begin{figure}
\begin{center}
\subfigure[\label{fig5a}]{\medfig{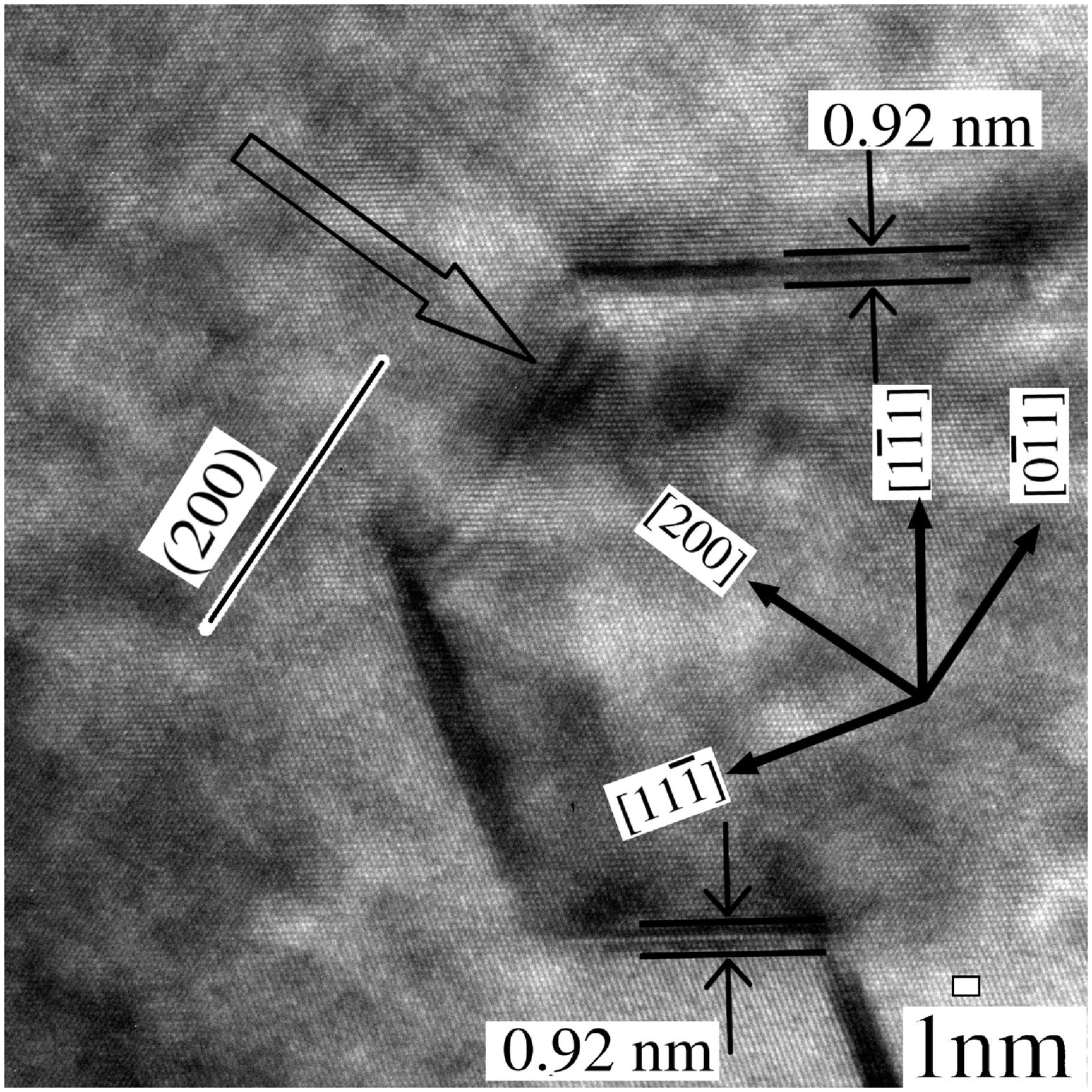}} 
\hfill
\subfigure[\label{fig5b}]{\medfig{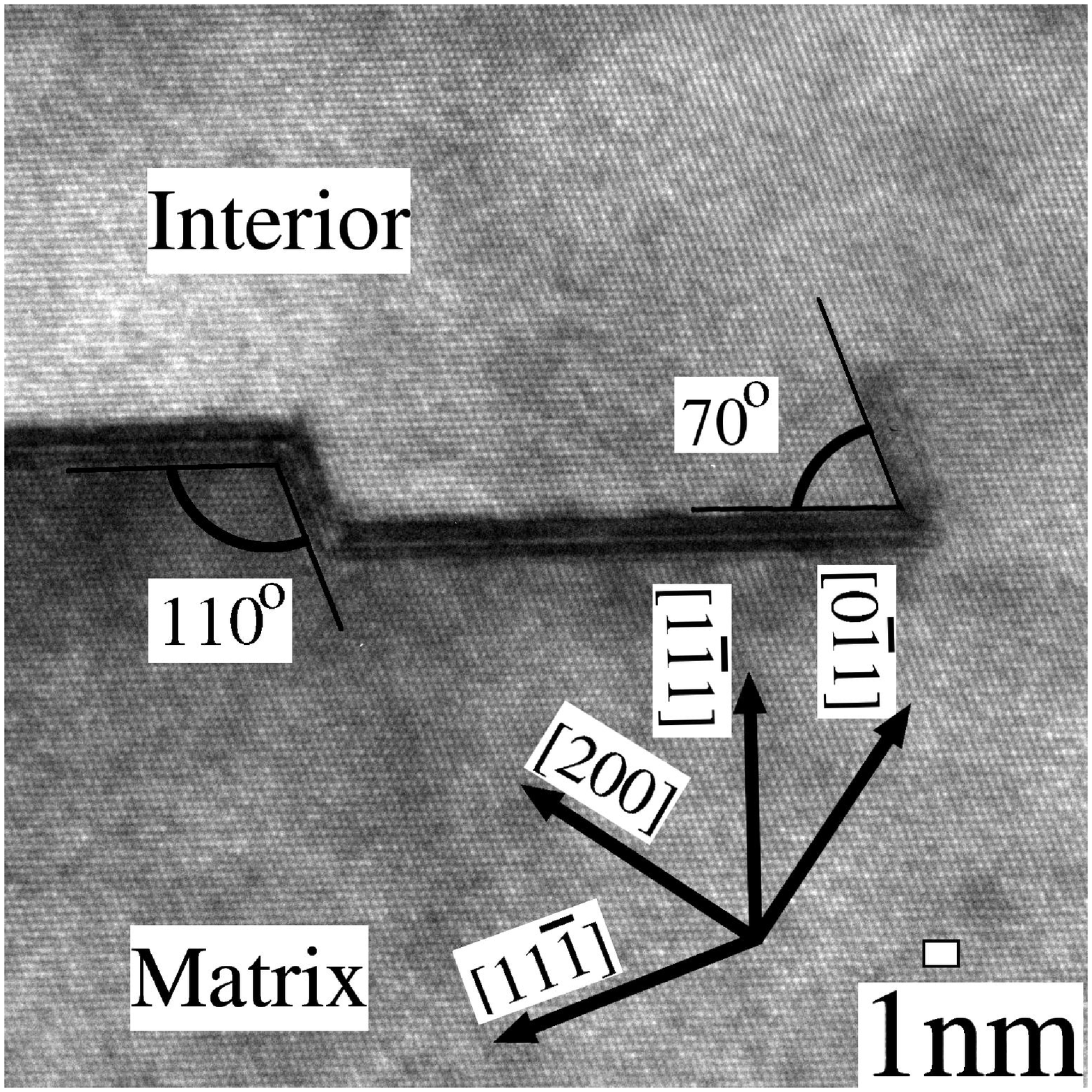}} 
\caption{High-resolution TEM micrographs of \alagcuB aged for 0.5\,h at 200$^\circ$C. 
Each image shows the end portion of a two-dimensional assembly structure of \gammaprime precipitates.  
A diffuse strain contrast is visible in (a) at the unoccupied section parallel to the (200) plane. 
In (b) the interior of the loop is towards the top of the field, with the matrix below. 
A single precipitate (right) is growing at an acute angle of 70$^\circ$ into the centre of the loop, in contrast to the more usual obtuse angle of 110$^\circ$ between adjacent precipitates. 
The precipitate thickness in both micrographs was measured and found to be 0.92\,nm, or two unit cells of the \gammaprime structure.\label{fig5}}
\end{center}
\end{figure}

High-resolution TEM was used to examine the structure of the precipitate assemblies after 0.016-0.5\,h ageing. 
The precipitates were extremely fine in scale, with diameters as low as \(5-6\)\,nm
and frequently, thicknesses of 0.92\,nm, 
equivalent to two unit cell heights of the \gammaprime phase.  
Figure~\ref{fig5a} 
 shows one example of the distal edge of a precipitate assembly. 
The diffuse contrast feature (see open arrow) noted in Figure~\ref{fig4a} 
(above) is clearly visible. 
The edge of another precipitate assembly is shown in Figure~\ref{fig5b}. 
The bottom of the field corresponds to the matrix, 
while the upper region was inside the assembly of \gammaprime plates. 
Note that the \gammaprime precipitate to the far right of the field is growing towards the interior of the assembly at an angle of 70$^\circ$ to the adjacent precipitate, rather than 110$^\circ$ that was observed at all other points around the assembly.

\subsection*{Precipitation of the $\boldsymbol{\thetaprime (\ce{Al2Cu})}$ phase}
\begin{figure}
\begin{center}
\subfigure[\label{fig6a}]{\stdfig{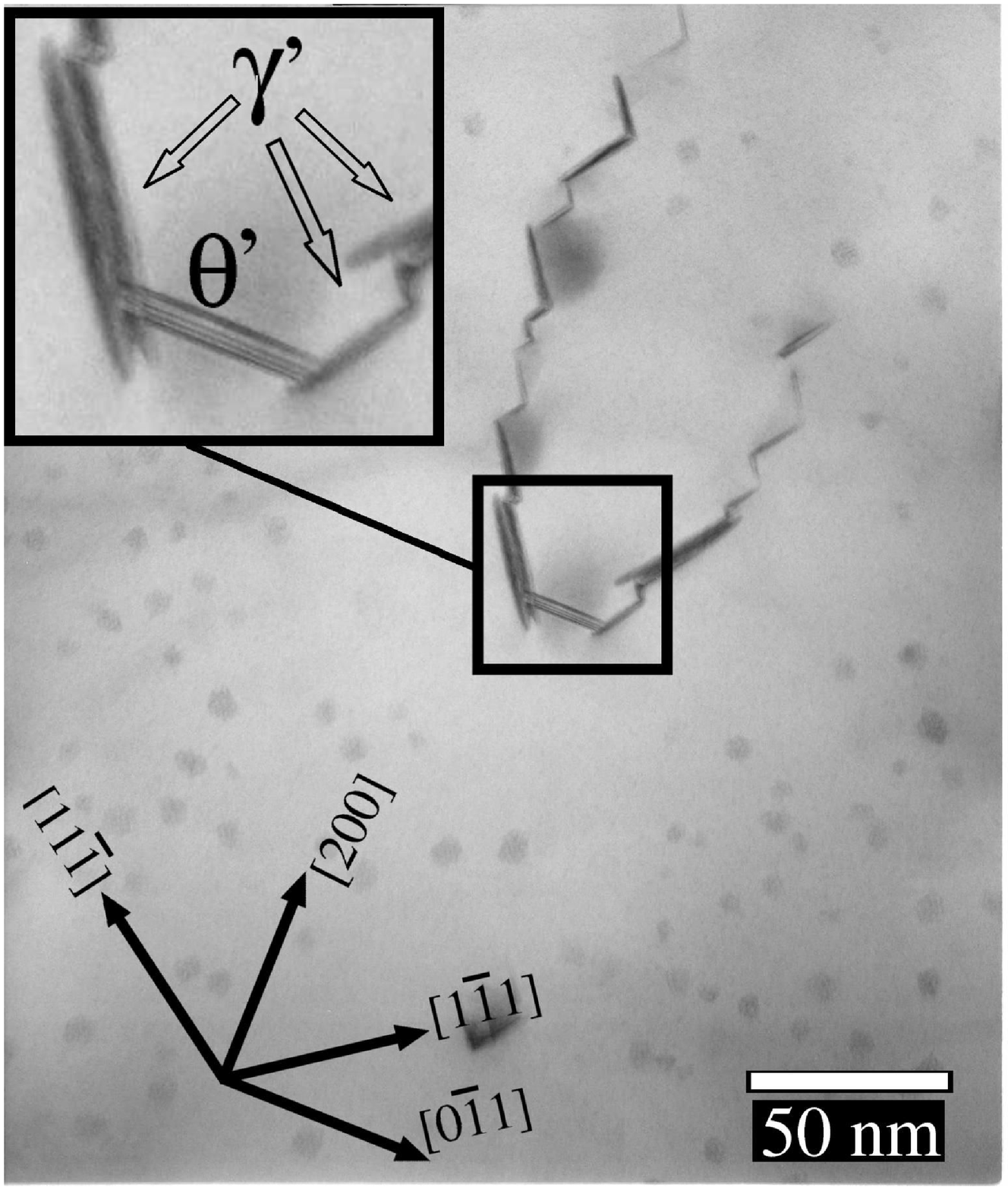}} 
\hfill
\subfigure[\label{fig6b}]{\stdfig{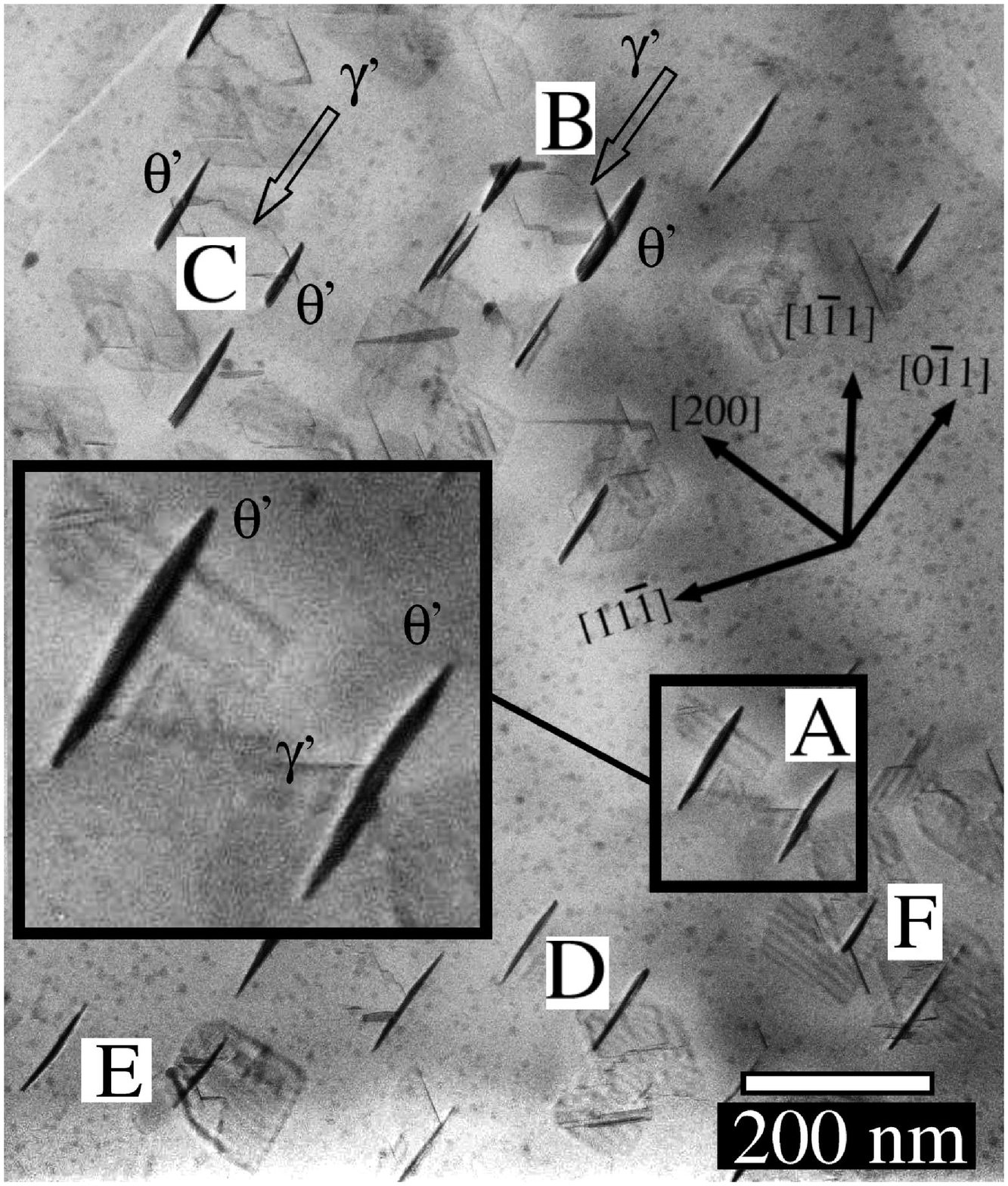}} 
\caption{\label{fig6}Electron micrographs showing the microstructure of the \alagcuB alloy after further ageing.
(a) After ageing for 2\,h at 200$^\circ$C. 
The micrograph shows precipitate assemblies containing both \gammaprime and \thetaprime precipitates. 
The \thetaprime precipitates have formed on the  \plane{200} section of the dislocation loop  and appear to be in direct contact with the pre-existing \gammaprime plates. 
(b) After ageing for 8\,h at 200$^\circ$C. The \thetaprime plates are prominent and the \gammaprime plate no longer form a continuous circuit.
Both images were obtained close to the \axis{011} zone axis.}
\end{center}
\end{figure}

The copper-rich \thetaprime (\ce{Al2Cu})
phase was first observed after 2\,h of ageing at 200$^\circ$C. 
This phase formed as thicker plates occupying the \planes{200} faces of the faceted elliptical precipitate assemblies
(See Figure~\ref{fig6a}).
Most precipitate assemblies were observed with \thetaprime plates on either one or both sides of the precipitate assembly, 
and in the latter case effectively formed a closed, two-dimensional  structure, as shown in Figure~\ref{fig6a}.

Further ageing results in the loss of the elliptical assembly structures and the continued growth of the \thetaprime phase.
A typical example of the microstructure in this stage is shown in Figure~\ref{fig6b}, 
where the alloy has been aged for 8\,hours. 
The \gammaprime precipitates were less prominent at this stage and the assembly structures have decomposed. 
The \gammaprime plates are most commonly observed in contact with \thetaprime precipitates,  towards either end of the remnants of the assembly structures. 
Guinier~-Preston zones can be seen in Figures~\ref{fig6a} and ~\ref{fig6b},
although the region close to the precipitate assembly in Figure~\ref{fig6a} 
is noticeably deficient in such zones.

\section*{Discussion}

The microstructure of the Al-Ag-Cu alloy contained both \gammaprime (\ce{AlAg2}) 
and \thetaprime (\ce{Al2Cu}) 
precipitates, 
in addition to spheroidal Ag-rich Guinier-Preston zones.
The \gammaprime phase formed at an earlier stage of ageing than the \thetaprime phase and adopted a very distinctive morphology where two variants of the phase alternated around an approximately elliptical aggregate. 
This was later capped at the sections tangential to the \plane{200} ends by the \thetaprime precipitates. 
It appeared that the \gammaprime precipitates grew more extensively in the plane of the loop, 
as is the case in Figure~\ref{fig4}.

\begin{figure}
\subfigure[\label{fig7a}]{\stdfig{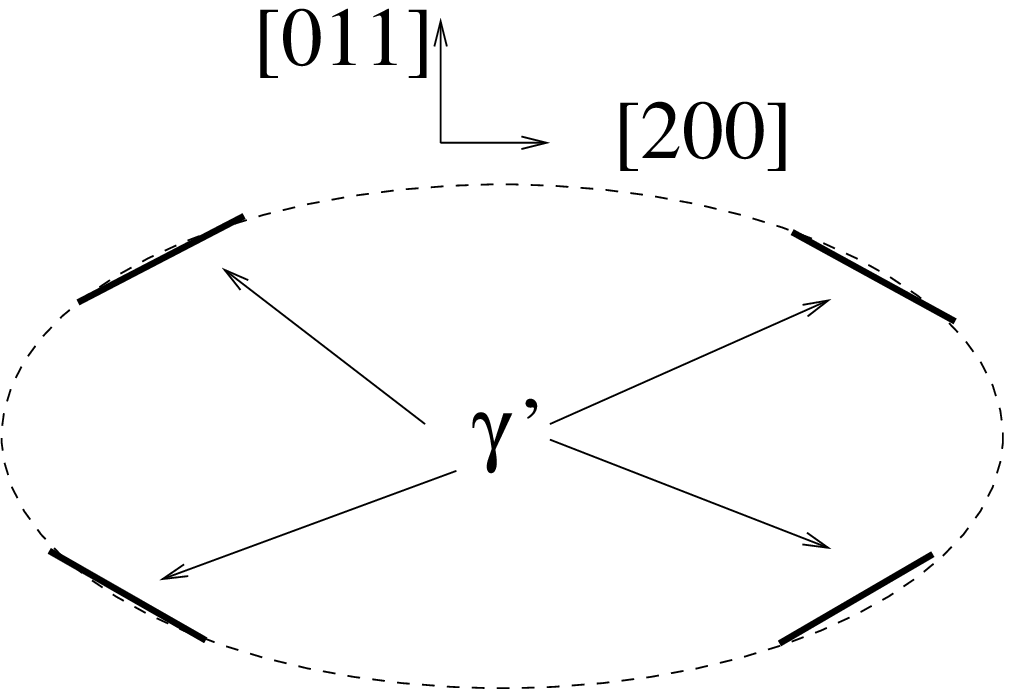}} 
\hfill
\subfigure[\label{fig7b}]{\stdfig{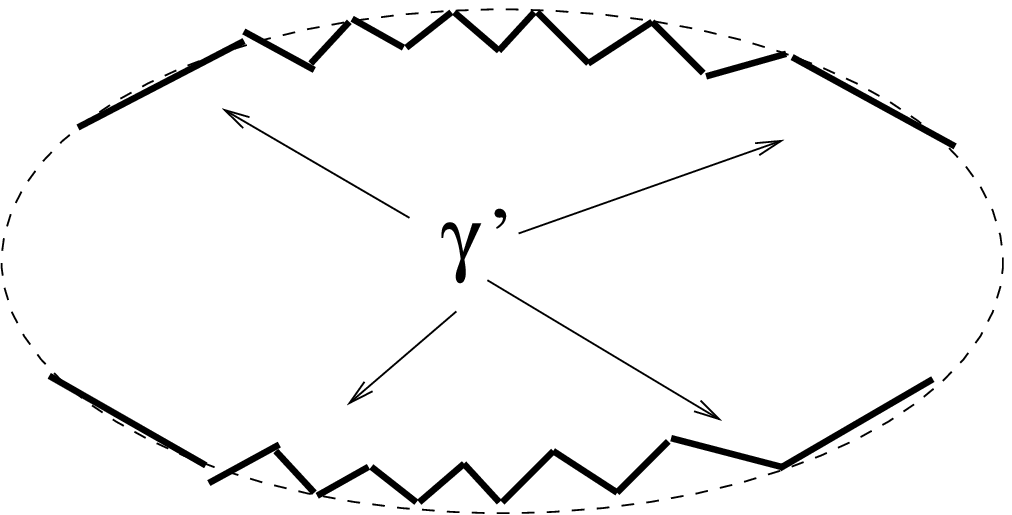}} 

\subfigure[\label{fig7c}]{\stdfig{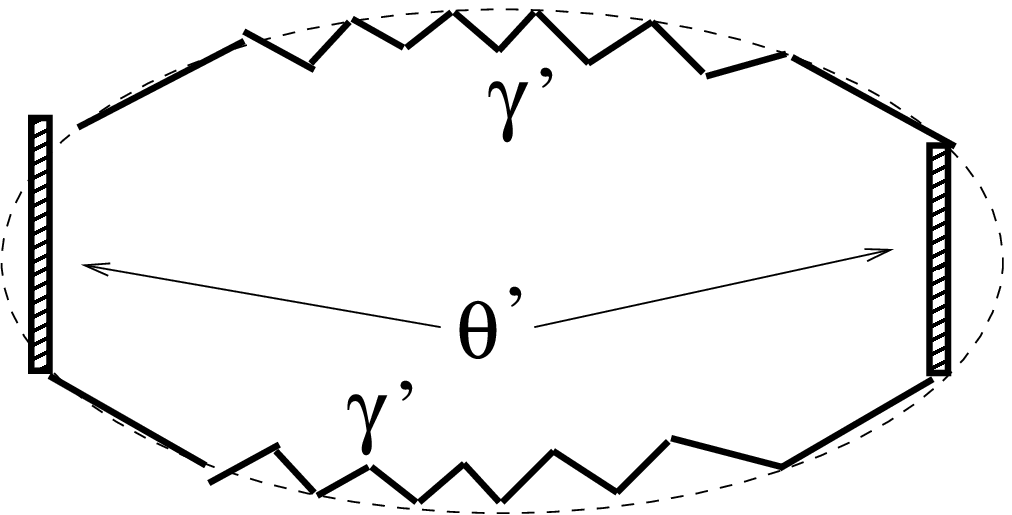}} 
\hfill
\subfigure[\label{fig7d}]{\stdfig{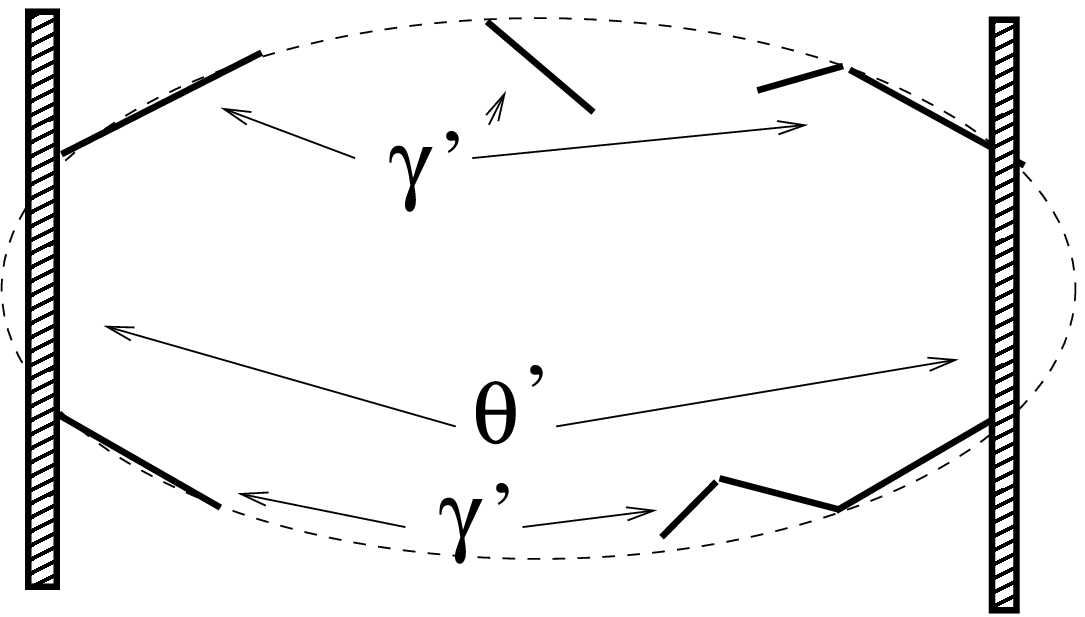}}
\caption{A schematic diagram illustrating the evolution of the 
\gammaprime-\thetaprime precipitate assemblies. 
(a) Precipitation of \gammaprime plates occurs on an elliptical perfect dislocation loop where the dislocation line is tangential to the habit plane.
(b) Further \gammaprime precipitates form, producing the faceted elliptical structure comprised of alternating variants of the phase.
(c) The \thetaprime precipitate nucleates at the segments of the loop tangential to the \planes[]{100} planes.
(d) Ostwald ripening results in the growth of the larger \gammaprime precipitates, at the expense of the smaller, later forming plates. 
The \thetaprime precipitates continue to lengthen.
\label{fig7}}
\end{figure}

The faceted elliptical assemblies have not been reported previously in the aluminium-silver-copper alloy system. 
The development of these precipitate assemblies is illustrated schematically in Figure~\ref{fig7}.
Precipitates formed furthest from the centre of the assembly (close to the \planes{200} normals), were typically of greater diameter and appear to have formed first (Figure~\ref{fig7a}).  
These were followed by alternating \plane{111} variants of the \gammaprime phase (b) and eventually \thetaprime precipitates at the \planes{200} ends of the assemblies (c). 
The assemblies later deteriorated due to coarsening of the precipitates, 
with the loss of the majority of the \gammaprime plates and the growth of the surviving \gammaprime plates and the \thetaprime precipitates (d). 
The presence of such assembly  structures can be explained through an examination of the interaction of the dislocations loops present in the as-quenched condition with the precipitate.

The dislocation loops were identified as the principle precipitation site of the \gammaprime phase. 
The precipitate assemblies are of similar scale to the dislocation loops (See Figure~\ref{fig3}) and have the same \planes[]{011} habit. 
Both dislocation loops and precipitate assemblies were absent from the grain boundaries, 
which are expected to act as vacancy sinks and thus prevent the formation of vacancy loops. 

\subsection*{Dislocation loops in Al-Ag-Cu alloys}

The dislocation loops in this alloy differed from those of binary aluminium-silver alloys\cite{westmacott:1971,Rosalie2008a}.
Dislocation loops in the \alagcuB alloy took the form of ellipses lying on the \planes[]{011} planes of the aluminium matrix. 
These did not display the alternate light-dark contrast which would indicate the presence of a stacking fault.
Dislocation loops in binary Al-Ag alloys had \planes[]{111} habit, 
were hexagonal in shape\cite{westmacott:1971,frank:1961} and  
displayed stacking fault contrast.

The difference in the character of the dislocation loops was ascribed to the differing effects of solute silver and copper on the matrix.
Hexagonal Frank dislocation loops have been  observed in Al-Ag alloy for compositions up to 20wt.\%Ag\cite{westmacott:1971}. 
In contrast, levels of copper as low as 0.2wt\% have been shown to alter the quenched in defect structure\cite{westmacott:1971}. 

The addition of copper to the aluminium-silver binary alloy is expected to affect both the  misfit strain energy and stacking fault energy. 
Copper generates considerable misfit strains in the aluminium matrix, 
due to the lower effective atomic size \cite{axon:1948}. 
 TEM studies of dislocation loops in aluminium-copper alloys found that dislocation loops did not exhibit stacking fault contrast\cite{westmacott:1971,thomas:1959}.  
The calculations of Schulthess \textit{et al}\cite{schulthess:1998} predict that the introduction of copper will increase the stacking fault energy of the aluminium matrix. 
Both factors will make the hexagonal Frank dislocations unstable and subject to further changes, 
commencing with the nucleation of a Shockley partial dislocation in the plane of the loop to remove the stacking fault\cite{hirsch:1958}.

Progressively greater additions of copper to aluminium alloys resulted in the formation of circular (undissociated) Frank loops, 
followed by (rounded) 
rhomboidal perfect loops on \planes{110} planes
and finally circular perfect dislocation loops, 
also on the  \planes[]{011} planes\cite{westmacott:1971}.
The elliptical dislocation loops observed in the present work correspond to the ``rounded'' rhomboidal dislocation loops reported by\cite{westmacott:1971}, 
as both appear to be perfect dislocation loops, 
elongated in the \axes{200} directions and 
lying on the \planes{011} planes of the matrix.
The net effect of the copper will therefore be to yield a different dislocation loop type, 
in which the dissociated Frank loops are replaced by elliptical perfect dislocation loops lying on \planes{011} planes.

The change in dislocation loop character is a complex process involving a number of steps, 
including the removal of the stacking fault via nucleation of a Shockley partial\cite{hirsch:1958}, glide of individual segments of the dislocation\cite{bullough:1964,bacon:1970} and the loss of two segments of the loop\cite{bullough:1964}. The stages preceding the formation of the elliptical dislocation loops have been  reviewed elsewhere\cite{rosalie:thesis}.

\subsection*{Precipitation of the \gammaprime phase on dislocation loops}

\begin{figure}
\begin{center}
\stdfig{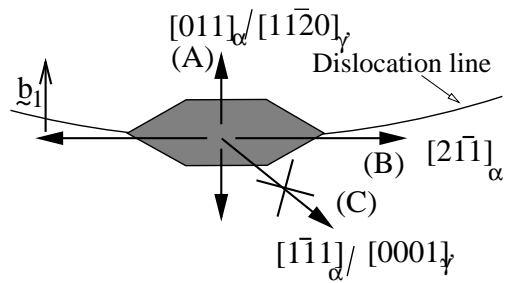}
\caption[The dissociation of a segment of an elliptical perfect dislocation loop to yield a stacking fault. ]
{The dissociation of a segment of an elliptical perfect dislocation loop to yield a stacking fault. 
(a) The perfect dislocation line with Burgers vector $\boldsymbol{b_1} = \mathrm{\frac{a}{2}[ 011 ] }$ has dissociated to form 
Shockley partial dislocations $\boldsymbol{b_2}$ and $\boldsymbol{b_3}$, with the  stacking fault between the partial dislocations indicated by the hashed region. 
The faulted region lies on the \plane{1\overline{1}1} plane and provides a nucleation site for the \gammaprime phase. 
\label{fig8}}
\end{center}
\end{figure}

Previous reports note that the \gammaprime phase can form as narrow rods on perfect dislocation helices\cite{frank:1961,nicholson:1961,passoja:1971}. 
Although the precipitate morphology is distinctly different from the assemblies described here, 
there are similarities since both nucleation sites possess Burgers vectors of \burgers{1}{2}{011} type.
For a dislocation helix, 
segregation of silver to the dislocation line resulted in the dissociation of segments of the 
\burgers{1}{2}{011}dislocation on \planes{111} planes to form two Shockley partial dislocations with 
Burgers vectors of type \burgers{1}{6}{112} (Figure~\ref{fig8}).
The hexagonal close-packed region between the two partial dislocations was able to act as a preferred site for the formation of the \gammaprime phase. 
The width \((A)\) of the precipitate parallel (in the \axis{011} direction in the figure) was controlled by the separation of the Shockley partial dislocations, 
which was determined by the stacking fault energy\cite{passoja:1971}. 
Lengthening \((B)\) in the \axis{211} direction required the dislocation line to straighten into the habit plane, and thickening (C) required the nucleation of additional ledges on the face of the precipitate. 

Precipitates formed on dislocation helices were restricted in length to the pitch of the helix\cite{passoja:1971}.
This was ascribed to the interaction between the dislocation helix and the precipitate.
The growth of the precipitate required the bowing of the dislocation segment into the operative
\plane[]{111}-type plane in order to allow dissociation to take place. 
This, 
in turn, 
required local adjustments in the shape of the helix 
(via dislocation glide and/or climb) 
which became more difficult to achieve as the precipitate increased in scale, 
limiting the length of the rod-like precipitates, 
giving rise to the series of parallel, 
rod-like \gammaprime precipitates along the helix.

A similar process of dislocation bowing may explain the formation of faceted elliptical assemblies on perfect dislocation loops in the present work. 
Silver segregation to the dislocation would result in the dissociation of the perfect dislocation, providing a potential nucleation site for the \gammaprime phase, as for precipitation on a dislocation helix. 
The width of the precipitate would also be controlled by the local stacking fault energy, 
resulting in more extensive growth in the plane of the assembly (as can be seen in Figure~\ref{fig4b}).

\begin{figure}
\begin{center}
\subfigure[]{\resizebox{4.5cm}{!}{\includegraphics{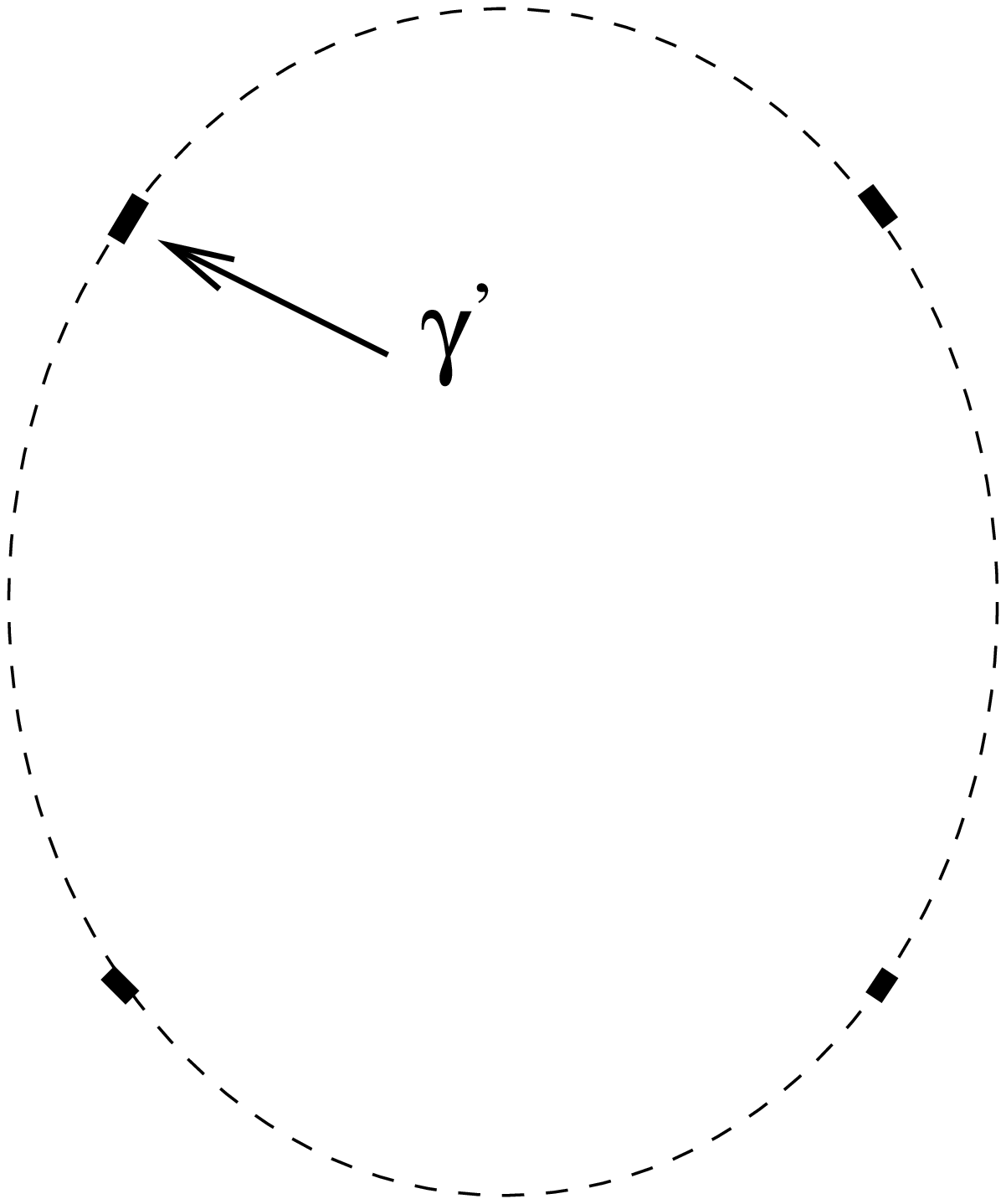}}}
\subfigure[]{\resizebox{4.5cm}{!}{\includegraphics{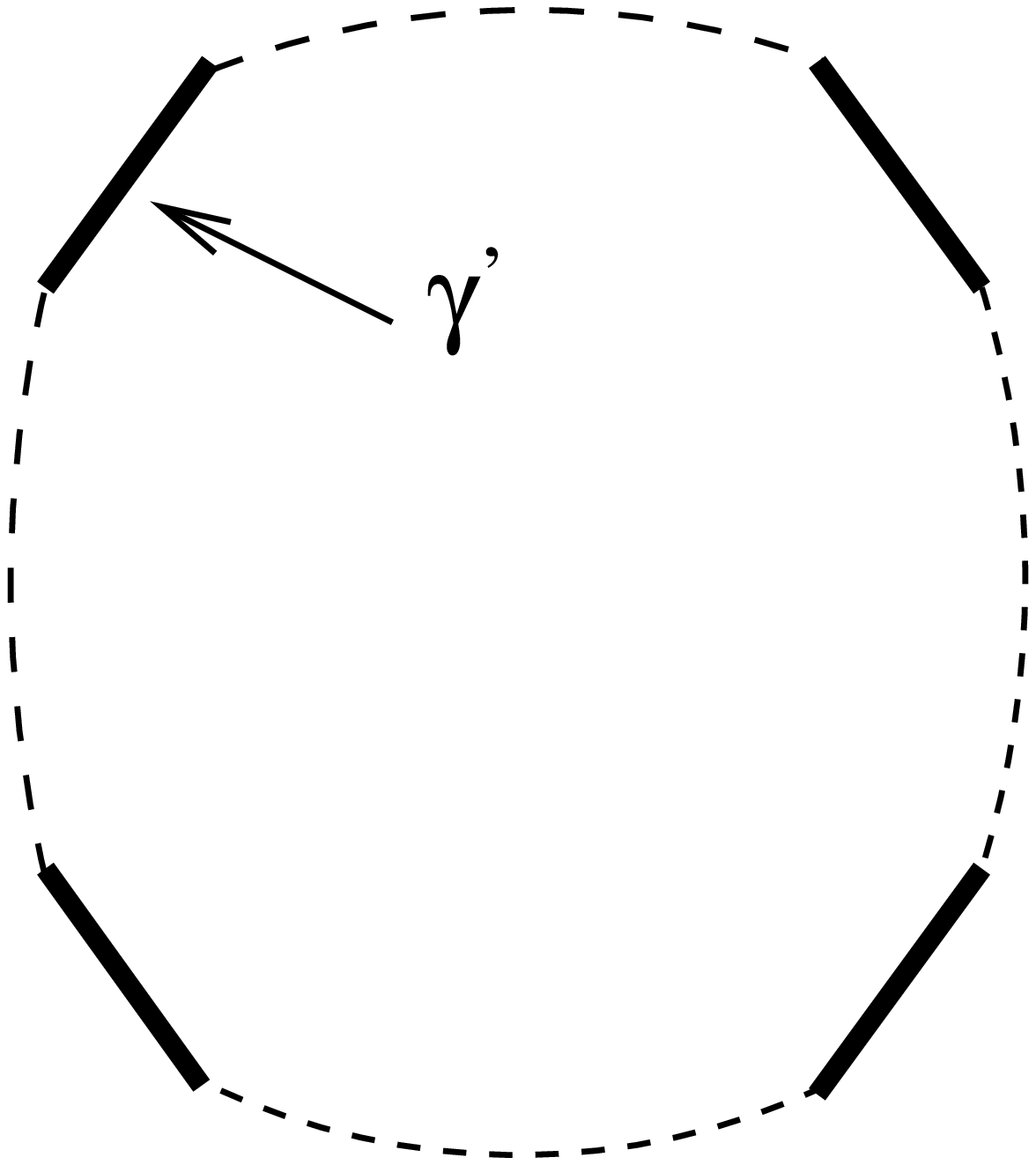}}}
\subfigure[]{\resizebox{4.5cm}{!}{\includegraphics{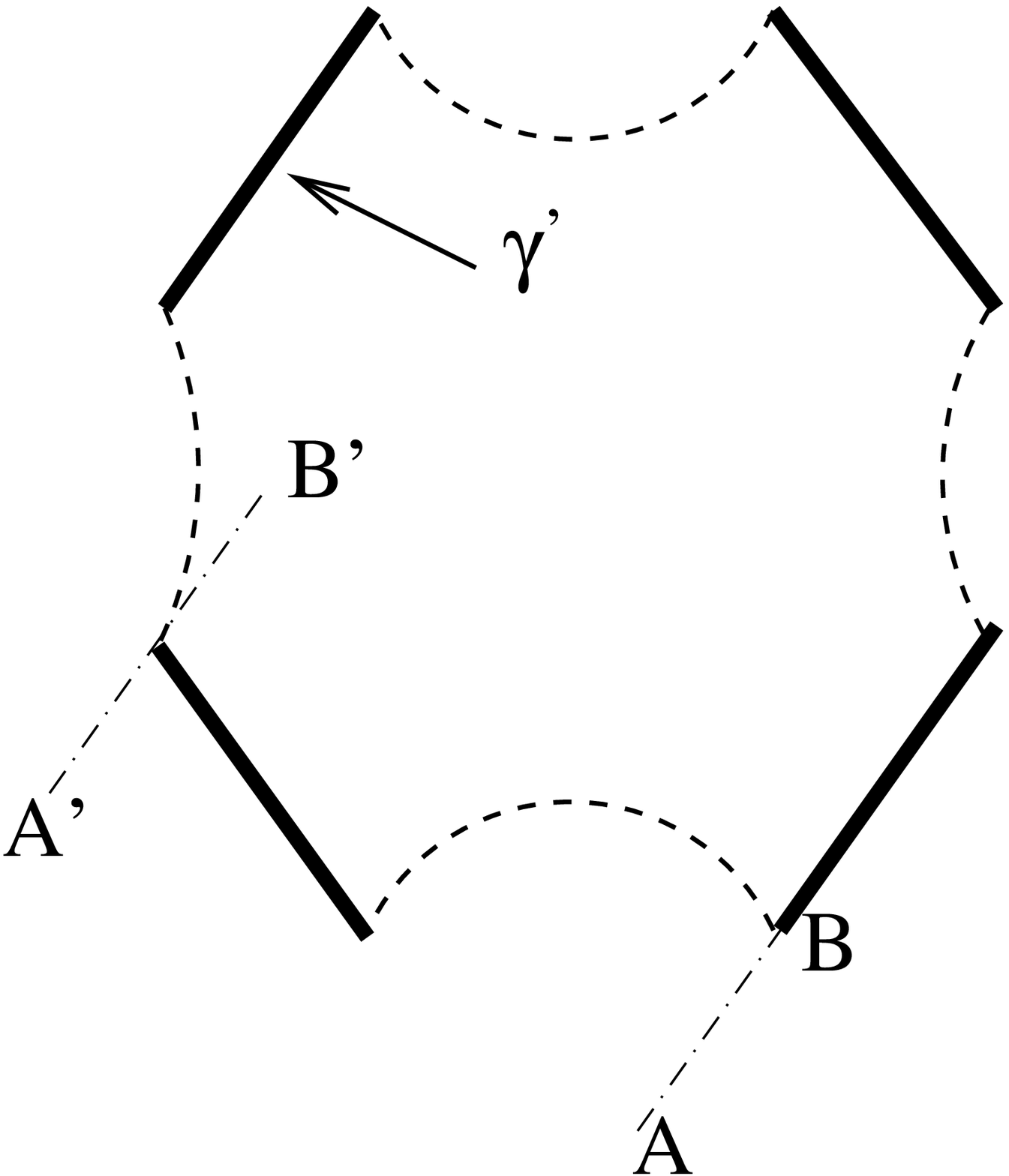}}}
\caption{A schematic diagram of the lengthening of the \gammaprime precipitates and the bowing of the dislocation loop. 
\gammaprime precipitates nucleated on the dislocation loop (in (a)),
lengthen, causing a flattening of the dislocation loop (in (b)).
If the precipitates lengthen sufficiently (as in (c)),
segments of the dislocation loop will become parallel to the alternate \planes{111} habit plane for the \gammaprime phase. (e.g. the loop tangent along \(A\prime B\prime\) has become parallel to one of the original precipitates along \(AB\)). 
This would offer the possibility of additional \gammaprime precipitates forming at this site, resulting in the faceted loop structure. 
\label{fig9}}
\end{center}
\end{figure}

Lengthening of the precipitate in the plane of the loop would lead to bowing of the dislocation loop, 
since dissociation of the perfect dislocation into Shockley partial dislocations must occur on a \plane{111} plane.
The dissociation of a segment of the loop along a \plane{111} plane would act to increase the area of the loop, 
unless accompanied by a contrary bowing of the loop to accommodate this change. 
This is illustrated schematically in Figure~\ref{fig9}. 
This figure shows \gammaprime precipitates nucleating on the dislocation loop in (a). 
These lengthen in the plane of the loop (in (b)), forcing the dislocation loop to bow inwards. 
Continued growth of these precipitates (c) would make segments of the dislocation loop parallel to the alternate \planes{111} habit plane for the \gammaprime phase. 
This may allow for the nucleation of  additional \gammaprime precipitates athese sites, resulting in the faceted loop structure reported here.

The presence of only two of the four \planes{111} variants of the \gammaprime phase in a given precipitate assembly was also due to the Burgers vector of the dislocation. 
A perfect dislocation of any given $\boldsymbol{b}=\frac{1}{2}\langle110\rangle$ vector can only dissociate on one of two \plane[]{111}-type planes, which are normal to the Burgers vector. 
These correspond to the two variants formed in each assembly, with habit planes normal to that plane of the faceted ellipse. 

\subsection*{Consequences of the precipitation of the \gammaprime phase in faceted elliptical assemblies}

Previous reports have noted qualitatively\cite{borchers:1969a}  
that the number density of \gammaprime precipitate was greater in the binary alloy compared to the ternary, 
and that the precipitates diameter was lower in the ternary alloy. 
More recent studies have concluded that the number density of \gammaprime precipitates in \alagcuB was an order of magnitude greater than for an \alagA alloy aged for equivalent time periods\cite{rosalie:2005}.

The higher number density of \gammaprime precipitates in Al-Ag-Cu alloys is a consequence of precipitation occurring on elliptical perfect dislocations, rather than hexagonal Frank dislocations. 
While the number of precipitates which can form on a hexagonal (dissociated) Frank loop is limited to the number of faulted surfaces initially provided by the dislocation loop\cite{Rosalie2008a}, 
the same restriction does not apply for elliptical dislocation loops. 
Lengthening of precipitates on a perfect dislocation loop can divert the dislocation line and result in the generation of \textit{additional} faulted surfaces on which the precipitate can form.
The number of precipitates per dislocation loop is significantly greater (see for example Figure~\ref{fig4a}, where 12 \gammaprime precipitates are visible).

The finer scale of precipitation can also be attributed to precipitation on the elliptical dislocation loop. 
The lengthening behaviour of the precipitates is markedly different in this alloy. 
In the binary alloy, 
lengthening was restricted by the nature of the dislocation bounding the stacking fault, 
with stair-rod \burgers{1}{6}{110} dislocations acting as barriers to growth, while growth was permitted where the stacking fault was bounded by Shockley partial dislocations\cite{Rosalie2008a}. 

For the faceted elliptical assemblies, 
the controlling factor for growth appears to be the potential for the nucleation of alternate \gammaprime variants due to the bowing of the dislocation loop. 
Nucleation of additional precipitates, 
of the alternative variant possible for a given dislocation loop,
would block the growth of the original plate in the plane of the defect. 

The  higher number density was not matched by a commensurate improvement in the mechanical properties of the ternary alloy.
Similar hardness data has been reported in the literature\cite{bouvy:1965},
but the reasons have not been made clear. 
The poor strengthening response was due in part to the localisation of the precipitates into discrete, isolated assemblies. 

The second factor limiting precipitation strengthening in the \alagcuB alloy relates to the limited growth of the precipitates. 
Plate-shaped precipitates offer optimal strengthening when thin and of extremely high aspect ratio\cite{nie:1996}. 
Growth of precipitates on elliptical, 
perfect dislocation loops does not allow the precipitates to adopt such a form and the precipitates could be more accurately described as lath-like.

\bigskip

The reduction in precipitate diameter and then increase in number density of
the \gammaprime (\ce{AlAg2}) phase in Al-Ag-Cu alloys compared to binary Al-Ag alloys 
can be attributed to the heterogeneous defect assisting in the nucleation of the precipitate. 
Nucleation on elliptical perfect dislocation loops in the ternary alloy
permits the formation of a greater number of precipitates per dislocation loop, 
while simultaneously restricting the lengthening of the precipitates.
Both effects are due to the lengthening of the precipitates causing bowing of the dislocation line and permitting the formation of additional \gammaprime precipitates, while simultaneously limiting the growth of those precipitates.
The \thetaprime precipitate was not observed until well after the faceted elliptical assemblies were well established and does not develop the extended cross- or T-shaped arrays reported in Al-Cu alloys \cite{perovic:1979,perovic:1980}.

While the presence of copper is responsible for the change in quenched-in defect structure,
change in microstructure is not due to the presence of the \thetaprime phase 
\textit{per~se}.

\section*{Conclusions}

A detailed examination of the microstructure of the ternary Al-Ag-Cu alloy has been conducted. 
The \gammaprime (\ce{AlAg2}) phase was found to form as faceted elliptical assemblies on perfect dislocation loops. 
Two variants of the \gammaprime phase were identified for a given dislocation loop, each lying on a \plane{111} normal to the habit plane of the dislocation loop. 
The formation of this structure can be explained through the dissociation of the dislocation loop, 
followed by bowing of the dislocation loop to compensate for the lengthening of the precipitate, leading eventually to the formation of the alternate \gammaprime variant.
The \thetaprime (\ce{Al2Cu}) phase was found to form on the dislocation loops following longer ageing and does not appear to be the determining factor leading to the change in microstructure.

The faceted assembly structure explains a number of features previously associated with the formation of the \gammaprime phase in the ternary alloy, 
namely the presence of a higher number density of the silver-rich precipitate, 
along with a more refined microstructure than in the binary Al-Ag alloy.
In both instances the precipitation-strengthening response is limited by the aggregation of the precipitates into discrete isolated assemblies.

The substantial differences in the precipitate microstructure in the binary Al-Ag and Al-Ag-Cu can be ascribed to differences in the heterogeneous defect assisting in the nucleation of the precipitate. 
This illustrates the importance of understanding the identity and role of heterogeneous defects in precipitation-strengthened alloys in general.

\section*{Acknowledgements}
JMR gratefully acknowledges receipt of an Australian Research Council APA scholarship and support through the ARC Centre of Excellence for Design in Light Metals.

\newpage

\end{document}